# Comparison of observed ground-motion attenuation for the 2016/04/16 $M_w$7.8 Ecuador megathrust earthquake and its two largest aftershocks with existing ground-motion prediction equations

C. Beauval[2], J. Marinière[2], A. Laurendeau[1], J-C. Singaucho[1], C. Viracucha[1], M. Vallée[3], E. Maufroy[2], D. Mercerat[4], H. Yepes[1], M. Ruiz[1], A. Alvarado[1]

[1] Instituto Geofísico, Escuela Politécnica Nacional, Quito, Ecuador

[2] ISTerre, Université Grenoble Alpes, IRD, CNRS, OSUG, Grenoble, France

[3] Institut de Physique du Globe, Paris

[4] CEREMA, Nice

Corresponding author:

Céline Beauval; celine.beauval@univ-grenoble-alpes.fr

**Abstract**

A megathrust subduction earthquake (Mw 7.8) struck the coast of Ecuador on April 16$^{th}$, 2016 at 23h58 UTC. This earthquake is one of the best-recorded megathrust events up to date. Besides the mainshock, two large aftershocks have been recorded on May 18$^{th}$, 2016, at 7h57 (Mw 6.7) and 16h46 (Mw 6.9). These data make a significant contribution for understanding the attenuation of ground motions in Ecuador. Peak ground accelerations and spectral accelerations are compared with four ground-motion prediction equations developed for interface earthquakes, the global Abrahamson et al. (2016) model, the Japanese equations Zhao et al. (2006) and Ghofrani and Atkinson (2014), and one Chilean equation Montalva et al. (2016). The four tested GMPEs are providing rather close predictions for the mainshock at distances up to 200km. However, our results show that high-frequency attenuation is greater for backarc sites, thus Zhao et al. (2006) and Montalva et al. (2016), which are not taking into account this difference, are not considered further. Residual analyses show that Ghofrani and Atkinson (2014) and Abrahamson et al. (2016) are well predicting the attenuation of ground motions for the mainshock. Comparisons of aftershock observations with Abrahamson et al. (2016) predictions indicate that the GMPE provide reasonable fit to the attenuation rates observed. The event terms of the Mw6.7 and Mw6.9 events are positive but within the expected scatter from worldwide similar earthquakes. The intra-event standard deviations are higher than the intraevent variability of the model, which is partly related to the poorly constrained $V_{S30}$ proxys. The Pedernales earthquake has produced a large sequence of aftershocks, with at least 9 events with magnitude higher or equal to 6.0. Important cities are located at short distances (20-30km) and magnitudes down to 6.0 must be included in seismic hazard studies. The next step will be to constitute a strong motion interface database and test the GMPEs with more quantitative methods.

# 1. Introduction

The megathrust Pedernales earthquake (Mw 7.8) struck the coast of Ecuador on April 16$^{th}$, 2016, at 23h58 UTC. Sixty-nine accelerometric stations recorded the earthquake at fault distances ranging from 26 to 427km (Fig. 1). One month after the mainshock, two large aftershocks have been recorded on May 18$^{th}$, 2016, at 7h57 and 16h46 (Table 1, Mw 6.7 and 6.9 respectively). The accelerometric network in Ecuador started in 2009 with 9 stations installed in the framework of the French-Ecuadorian research project ADN ("Andes du Nord"). In 2010, the Ecuadorian research agency SENESCYT granted the Geophysical Institute in Quito with an ambitious project for instrumenting the whole country with high-level instruments, accelerometric, broad-band and GPS stations. The accelerometric network, now called RENAC, is still in a developing phase, with ~30% of the stations telemetered, and the characterization of the sites undergoing.

Ecuador is exposed to a high seismic risk, both from earthquakes on the subduction interface like the 2016 event, and from earthquakes on shallow crustal faults in the Andean Cordillera. Since 2007, a French-Ecuadorian cooperation aims at leading research on all aspects related to probabilistic seismic hazard assessment (PSHA), in order to improve PSHA in Ecuador (e.g. Alvarado et al. 2014; Beauval et al. 2010, 2013, 2014; Yepes et al. 2016). PSHA aims at providing ground motions with probabilities of being exceeded in future time windows. The results can be used to establish seismic zoning for national building codes. As the strong motion database was still in its development phase, no study has been published yet on the testing of GMPEs against accelerometric data. In the PSHA calculations, GMPEs have been selected based on tectonic similarities criteria (e.g. Beauval et al. 2014). The Mw7.8 earthquake and its largest aftershocks have produced a unique dataset. These data make a significant contribution for understanding the attenuation of ground motions in Ecuador. In the present study, peak ground accelerations (PGA) and spectral accelerations are compared with four ground-motion prediction equations (GMPEs) developed for interface earthquakes, the recent global Abrahamson et al. (2016) model as well as two Japanese equations, Zhao et al. (2006) and Ghofrani and Atkinson (2014), and a new Chilean model Montalva et al. (2016).

## 2. Strong-motion data

Strong-motion data are obtained from the National Accelerometric Network (Red Nacional de Acelerógrafos, RENAC), which includes seven accelerometers of OCP (Oleoducto de Crudos Pesados) as well as nine ADN accelerometers. Fig. 1 shows the distribution of the 69 stations triggered by the Pedernales event relative to the earthquake fault plane surface projection. There are sixteen stations at rupture distances ranging from 26 to 100km, distributed in the coastal plain. Thirty-three stations are located in the North-South Andean Cordillera hosting many volcanoes, 14 stations are installed in the Quito basin (Laurendeau et al. 2017). Approximately half of the stations are located in the forearc region, west of the volcanic front, and the other half lies in the backarc region. The records at six example stations are displayed in Fig. 2. The Mw 6.9 and 6.7 aftershocks were recorded respectively by 61 and 64 stations; 5 of these stations did not record the mainshock (Fig. 1).

All stations are installed on the ground surface and record continuously. Different digital accelerometer devices are used, Guralp, Reftek and Kinemetrics (See Table 2). For this study, a simple processing was applied. Acceleration time histories were visually inspected and windows extracted. A first-order baseline operator and a simple baseline correction are applied on each window for each component. Signal-to-noise Fourier spectral ratios have been carefully calculated with the signal processing tools of Perron et al. 2016. Given the magnitude of the three events, these ratios are in most cases high for the frequencies of interest (PGA and 0.5 to 5 Hz). At the stations located at distances between 300 and 500km, the signal-to-noise ratios are still higher or equal to 3 in this frequency range. Response spectra were then calculated with critical damping 5%. For each record, the geometric-mean horizontal-component is calculated for peak ground acceleration and spectral periods up to 2s.

The site conditions at a recording station have a strong influence on ground motions. The most common proxy for the simplified classification of a site in terms of its seismic response is $V_{S30}$, the time-average shear wave velocity in the upper 30 m. In Ecuador, few RENAC stations have been characterized with geophysical methods and significant efforts still need to be made to evaluate the geotechnical information of the sites. In Quito (14 sites), $V_{S30}$ are inferred from geophysical investigations of the subway project (TRX - Consulting C.A, 2011) and from a microzoning study (ERN 2012). For each station, $V_{S30}$ is inferred from the shear-wave velocity profile closest to the site. In Guayaquil (3 sites), $V_{S30}$ values come from the work of Vera-Grunauer (2014). A new project was started after the mainshock by the Geophysical Institute to investigate the site effects in the coastal cities, which to this date yields $V_{S30}$ values for 3 sites based on MASW technics.

For the other sites, following Zhao et al. (2006), H/V response spectral ratios are computed to determine the natural period of the site (Tg) and classify the sites into 4 broad site classes (SC I, II, III, IV, from rock to soft soil). The number of recordings available at each station varies from 3 to 203 (15 on average, see Laurendeau et al. 2016). The entire signal windows are used. At 33 sites, the natural period can be estimated, and $V_{S30}$ is deduced as $V_{S30}=4H/Tg$ with H=30m. At 16 sites with a natural period estimated higher or equal to 0.6s (soft soil, SC IV, Zhao et al. 2006), $V_{S30}$ is fixed to 200 m/s. At 6 sites showing a flat H/V ratio with amplitudes lower than 2, the site is classified in the rock/stiff soil class with a $V_{S30}$ of 800 m/s. There are 14 sites for which there was no clear peak but broadband frequency amplification. The method cannot be applied and an average $V_{S30}$ of 400m/s is arbitrarily attributed. More work is required to understand the limits of the method and how to adapt it to sites in Ecuador. This set of estimated $V_{S30}$ is considered as the reference $V_{S30}$ set.

To take into account the huge uncertainty on the $V_{S30}$ values, a second set of $V_{S30}$ is used. It is based on the weighting of the four closest points given in the database on the Global USGS $V_{S30}$ Slope Topography web site (see section Data and resources). These $V_{S30}$ are based on a relationship between the topographic slope and $V_{S30}$ (Wald and Allen 2007). The $V_{S30}$ values based on topography are compared to the reference $V_{S30}$ values in Fig. S1 (available in the electronic supplement). At distances shorter than 100km, in the forearc region, the difference does not exceed 200m/s. At larger distances, up to 600m/s difference can be observed for stations in the Cordillera. In the present work, the comparisons between observations and predictions are systematically led for both $V_{S30}$ sets, showing that this uncertainty does not impact the results. All results displayed in the paper rely on the reference $V_{S30}$ set, while results based on the alternative $V_{S30}$ set based on topography are in the electronic supplement.

**3. GMPEs selected**

Ground motion prediction equations describe the median and the variability of ground-motion amplitudes, depending on magnitude, site-source distance, site conditions, and other parameters. Four equations are considered here, two Japanese models, one Chilean and one global model: Zhao et al. (2006), Ghofrani and Atkinson (2014), Montalva et al. (2016) and Abrahamson et al. (2016). Zhao et al. 2006 does not include the recent interface events but proved to be quite stable and to fit reasonably the data available in South America (e.g. Arango et al. 2012, Beauval et al. 2012). Abrahamson et al. (2016) model is our favorite candidate for PSHA applications, as it includes the largest amount of global data, and an earlier version of the model proved to be stable and well-fitting datasets from various subduction environments (Beauval et al. 2012). All four models use the geometric mean of the two horizontal components, moment magnitude and rupture distance (closest distance to the fault plane). All are providing the total sigma, as well as the intraevent (variability from the median predicted value for a particular recording station in a given earthquake) and interevent variabilities (variability between earthquakes of the same magnitude).

Abrahamson et al. (2016) is based on the combined datasets used in several of the past subduction GMPEs (e.g. Atkinson and Boore 2003 and Youngs et al. 1997), as well as additional ground motion data obtained in Japan, Taiwan, South and Central America and Mexico. This new global GMPE is intended to replace the older global GMPEs. The metadata was carefully checked and improved, and recent events around the world were included. The final dataset includes 43 interface earthquakes (6.0≤Mw≤8.4) at distances up to 300km. Fifty-seven percent of interface records are from Japan, twenty-nine percent from Taiwan. The model is predicting a stronger attenuation for sites located in the backarc region with respect to sites located in the forearc region. The model is including site nonlinearity.

Ghofrani and Atkinson (2014) developed a ground motion prediction equation for interface earthquakes of Mw 7.0 to 9.0 based on data from Japan. The >600 strong ground motions records from the Mw9.0 2011 Tohoku earthquake are used to derive an event-specific GMPE, which is then extended to represent the shaking from four other Mw>7.0 interface events in Japan which occurred in 2003, 2004 and 2005. Three GMPEs are finally available to represent the epistemic uncertainty, an upper and lower model, as well as a median model. The median central model is used here. The equation accounts for the difference in the attenuation between forearc and backarc region by using separate anelastic attenuation factors. The soil response is treated as linear. Ghofrani and Atkinson (2014), like Abrahamson et al. (2016) explicitly use $V_{S30}$.

Zhao et al. (2006) developed an attenuation model for Japan based on events with Mw 5.0 to 8.3, at distances up to 300km. Four site classes are used in the present study, SC I, II, III and IV, approximately corresponding to the four classes, rock, hard soil, medium soil, and soft soil (site classification scheme used in Japanese engineering design, Zhao et al. 2006). The authors associate to these site classes approximate NEHRP site classes and $V_{S30}$ intervals (Table 2 in Zhao et al. 2006). The near-source data (<30km) is mostly constrained by the records from crustal events, however this should not affect the predictions for subduction events for distances >30km.

Montalva et al. (2016) developed a ground motion prediction equation relying on Chilean subduction earthquakes that occurred between 1985 and 2015, including the three recent megathrust earthquakes (Maule 2010, $M_w$ 8.8; Iquique 2014, $M_w$ 8.1; Illapel 2015, $M_w$ 8.3). The median model is based on the same functional form as Abrahamson et al. (2016). The attenuation is predicted only for forearc sites, as all recording stations are located in the forearc region. Montalva et al. (2016) indicate that the number of strong-motion stations with measured $V_{S30}$ is limited and that $V_{S30}$ proxies have been inferred both from the topographic slope (Wald and Allen 2009) and the site's predominant period (Zhao et al. 2006). Montalva et al. (2016), like Abrahamson et al. (2016) and Zhao et al. (2006), do not include data beyond 300km.

Abrahamson et al. (2016) and Ghofrani and Atkinson (2014) predict different attenuation depending on the location of the station with respect to the volcanic front. The forearc region is between the subduction trench axis and the axis of volcanic front. The backarc region is behind the volcanic front. The high-attenuation, low-velocity region in the crust and upper mantle related to the volcanic activity filters the high-frequency content of ground motion, as shown by Ghofrani and Atkinson (2011) on inslab events and by Ghofrani and Atkinson (2014) on interface events.

Most of the interface models published up to now have been coded in the strong-motion toolkit used here for predicting accelerations (Weatherhill 2014). This toolkit relies on the GMPE libraries of the OpenQuake PSHA software (Pagani et al. 2014). The Lin and Lee (2008) GMPE established on Taiwanese data was not selected as the equation is using the hypocentral distance, and given the short distances involved in Ecuador this might not be adequate. The Kanno et al. (2006) GMPE is not included because it would be a third Japanese model and it uses an unconventional definition for the horizontal component of motion. The Mexican equation by Arroyo et al. (2010) is not considered either because it predicts ground motions at rock sites only (NEHRP B class).

## 4. Fault plane solution and distance calculation

The site-source distances are calculated using the closest distance to the fault rupture plane (rupture distance). The fault must be approximated by a rectangular plane. There is no unique solution for the finite-fault plane (e.g. Goda and Atkinson 2014). Different fault models can be derived using various datasets and methods in source inversion analysis. The inversion might include GPS data, InSAR, teleseismic body wave, surface wave data, and near-source strong motion data. Goda and Atkinson (2014) explored the uncertainty related to the choice of the rupture plane for three Japanese mega-thrust earthquakes and showed that the impact on the comparison between observations and models can be significant. For now, for the 2016 Pedernales event, we are aware of only one elaborated model by Nocquet et al. (2016). The maximum slip is about 6.2m. From this slip model, we extracted the fault plane which includes approximately the 100 cm slip contour. The resulting plane is a rectangular of 100 km in length and 50 km in width, dipping to the East with a strike of 26.50° and a dip of 23°, extending from 13 to 33 km (Table 1, Fig. 1). The hypocenter solution is updip on the northern border of the fault plane.

The Pedernales earthquake is one of the best-recorded megathrust events up to date, in terms of distribution of stations around the fault plane and number of recording stations. Records are available above the fault plane (2 stations, Fig. 1), at short distances from the fault plane to the North and Northeast (10 stations between 45 and 100km), East (2 stations at 73 and 103km), South and Southeast (4 stations between 40km and 75km).

The rupture distance measure, taking into account the extension of the fault plane, only captures macroscopic features of the source. The more detailed components of recorded strong motions in the near-source region are not taken into account (e.g. short periods affected by local asperities). Besides, the 2016 Pedernales event is presenting evidences of directivity effects, with higher ground motions in the direction of the slip, south of the rupture plane, than in the north. These observations cannot be modeled by current published interface GMPEs.

As no fault plane solution has been inverted yet for the aftershocks, the length and width of the faults are based on Strasser et al. (2010) relations (Table 1). The fault plane is arbitrarily centered on the hypocenter.

## 5. Comparing observations and predictions

### 5.1 Mainshock Mw7.8

At first, predictions and observations are compared based on simple attenuation plots. As a second step, residual analyses are performed where the predictions include the $V_{S30}$ for each site.

Predictions from Abrahamson et al. (2016) are superimposed to the observations, for the PGA (Fig. 4a). To begin with, predictions are provided with the "forearc/unknown" option (Abrahamson et al. 2016). Three $V_{S30}$ are considered (200, 400 and 760 m/s), producing slightly different amplitudes. The attenuation rate predicted is consistent with the observations for distances lower or equal to 130km. For distances between 130 and 400 km, the observed attenuation rate appears steeper than predicted. Stations within 130 km from the rupture plane are all in the forearc region. At distances larger than 130 km, half of the stations are within or behind the volcanic arc (Fig. 1). Taking into account the backarc option in the equation yields a steeper attenuation with distance, in accordance with the observations at backarc stations (Fig. 4c). In the capital Quito, located at around 150km from the earthquake in the Cordillera (Fig 1), recorded PGA varies between 0.017 and 0.081g.

The rupture propagated to the South, producing directivity effects on ground motions. At rupture distances 40 to 80km, stations located to the South of the rupture experienced larger amplitudes than stations located to the North or to the East (Figs. 1 and 5). A specific study will need to be performed to investigate the source contribution on the Pedernales ground motions. The recorded data might need to be corrected for path and site effects to explain the difference of amplitudes in terms of source directivity (see e.g. Cultrera et al. 2009).

As expected, long-period ground motions decay less rapidly with distance than do short period motions. Fig. 4b displays predictions superimposed to observations at T=1.0s. Amplitudes predicted are more $V_{S30}$-dependent than for short period. Overall, the attenuation rate predicted is consistent with observations. Considering predictions for $V_{S30}$ from 200 to 760 m/s and considering the predicted variability (total sigma), most of the observations are within the predicted range. The model predicts similar decay with distance for forearc and backarc regions, and the observations indeed do not present significant differences (Fig. 4d).

Predictions by Zhao et al. (2006), Ghofrani and Atkinson (2014) and Montalva et al. (2016) are now superimposed to the observed data, considering an average $V_{S30}$ value (400m/s, Fig. 6). For distances in the range 30-150km, PGA median predictions from the four GMPEs are quite similar and consistent with the observed attenuation rate, with around 0.3g predicted at 40km and 0.1-0.11g at 100km. The total sigmas predicted are also close. For distances larger than 150km, the two Japanese models predict stronger distance decay. Zhao et al. (2006) does not differentiate attenuation between forearc and backarc stations, but its generating dataset includes many Japanese backarc stations. Applying the forearc/backarc station classification, Ghofrani and Atkinson (2014) predicts a stronger attenuation for backarc stations at distances larger than 100 km, with predictions very close to Abrahamson et al. (2016) (Fig. 7a, PGA). At T=1.0 s, Abrahamson et al. (2016) predicts larger accelerations at distances < 200km than the Japanese and Chilean models (Fig. 6b). Note that the generating datasets of Abrahamson et al. (2016), Zhao et al. (2006) and Montalva et al. (2016) do not include records beyond 300km and the models are therefore extrapolated at these distances.

To more accurately evaluate the performance of the GMPEs relative to the data, total residuals are calculated considering $V_{S30}$ for each station ($V_{S30}$ reference set, see Section "Strong Motion Data"). Residuals are calculated first ignoring the forearc/backarc distinction, and then including this attenuation difference. At the PGA, a trend in the distance dependence of residuals is observed with backarc sites showing a negative slope (Fig. 8a). Applying backarc coefficient to the sites in the backarc region, the slope becomes flatter, with mean residuals closer to zero (Fig. 8b). The same observation can be made for the residuals relative to Ghofrani and Atkinson (2014) equation (Fig. S2, available in the electronic supplement). At 1 second, as expected, no difference can be seen in the distance-decay rates for the forearc and the backarc stations (Fig. 8c-d). Ghofrani and Atkinson (2014) is slightly under-estimating the observations, as shown by the mean residuals higher or equal to zero.

At present, the uncertainty on $V_{S30}$ estimate is huge for the RENAC stations (see Section "Strong Motion Data"). The second set of $V_{S30}$ values based on topographic slope is considered as an attempt to evaluate the impact of $V_{S30}$ uncertainty on the results. The residuals obtained with Abrahamson et al. (2016) model are presented in the Electronic Supplement (Fig. S3). Residuals are quite stable with respect to the previous ones. At the PGA, for distances lower than 100km, residuals are identical to the ones calculated with the reference $V_{S30}$ set. This is expected; at these distances the difference in the $V_{S30}$ values is not exceeding 200 m/s (Fig. S1). At larger distances, only slight difference in the mean residuals can be noticed. At spectral period T=1.0s, mean of residuals are slightly shifted to positive values with respect to Fig. 8, but still no major change is observed. Throughout the study, all residuals have been derived on both sets of $V_{S30}$ values, showing that the results are stable.

**5.2 Aftershocks Mw6.9 and Mw6.7**

As the mainshock data shows clearly an attenuation effect due to wave passage through the volcanic front, the models applied in Ecuador should take this difference into account. Ghofrani and Atkinson (2014) equation is made for events with magnitude higher than 7.0. Thus, only Abrahamson et al. (2016) model is considered further for the aftershocks.

Figure 9 shows geometric mean PGA and T=1.0s spectral acceleration as function of rupture distance for the Mw6.9 event. The median and sigma predicted by Abrahamson et al. (2016) model are superimposed to the data, for an average $V_{S30}$ value of 400m/s. The residuals are also calculated. Observations are more scattered than for the mainshock, however comparable observations can be made. The attenuation rate predicted is roughly consistent with the observations, with a stronger attenuation at backarc sites for PGA. Mean of residuals are in general within one standard deviation. At T=1.0s, mean residuals at distances larger than 150km are larger or equal to sigma, indicating that the model is predicting a stronger attenuation than observed.

Results for the Mw6.7 aftershocks are displayed in Fig. 10. At short period (PGA), the difference in attenuation between forearc and backarc stations is less clear (Fig. 10a). The attenuation rate for backarc sites appears to better fit the observations for distances larger than 100km for all stations (forearc and backarc). Residuals indeed show a negative slope (Fig. 10c). The residuals at T=1.0s show a flatter slope, with positive mean residuals at distances larger than 100km indicating that the model is predicting on average lower ground motions than observed (Fig. 10d). Part of the data is indeed above the predictions (Fig. 10b).

**5.3 Events terms and intra-event standard deviations**

For the three events, the event term and intra-event standard deviations are calculated for a suite of 6 periods between PGA and 2 seconds (Fig. 11). Residuals at distances larger than 300km, the validity limit of the Abrahamson et al. (2016) model, are not included. The event term is the mean of the residuals in a single event over all stations. The intra-event residual is the misfit between an individual observation at a station from the earthquake-specific median prediction, which is defined as the median prediction of the model plus the event term for the earthquake (Al Atik et al 2010). The general trend of the event terms with spectral period is consistent for the three earthquakes (Fig. 11a). Event terms are mostly within the expected scatter for interface subduction earthquakes worldwide ($\tau$=0.43). Event terms are both negative and positive for the mainshock, but always positive for the aftershocks (larger than expected ground motions). Intra-event standard deviations for the mainshock are close to the expected scatter ($\phi$=0.6) for spectral periods lower than 1 second (Fig. 11b). At 1 and 2 seconds, the intra-event variability is higher than expected. This might be partly due to the poorly constrained $V_{S30}$ parameter and to the directivity effects on the ground motions.

Residuals, event terms and intra-event standard deviations based on the second set of $V_{S30}$ values, relying on topography, are displayed in Figs. S4-5 (available in the electronic supplement). Results are quite stable with respect to the calculations based on the reference $V_{S30}$. Intra-event standard deviations are again higher or equal to the intra-event variability predicted by the model.

**Conclusions**

The Pedernales interface earthquake of April 16th, 2016, has produced a unique dataset which enable to analyze the attenuation of ground motion with distance in Ecuador, and to evaluate the performance of interface models currently in use to predict strong ground motions in seismic hazard studies. The national accelerometric network RENAC is young and most stations still require site characterization, limiting the precision in the comparison of observations with existing ground-motion models.

The four considered GMPEs, Zhao et al. (2006), Ghofrani and Atkinson (2014), Montalva et al. (2016) and Abrahamson et al. (2016) are providing rather close predictions for a Mw7.8 earthquake at distances up to 200km. However, our results show that high-frequency attenuation is greater in the backarc region, thus Zhao et al. (2006) and Montalva et al. (2016), which are not taking into account this difference, are not considered further. Overall, residual analyses show that Ghofrani and Atkinson (2014) and Abrahamson et al. (2016) are rather well predicting the attenuation of ground motions for the mainshock, both for short and long periods. A specific study investigating the signature of directivity effects in the recorded ground motions remains to be done.

Comparisons of aftershock observations with Abrahamson et al. (2016) predictions indicate that the GMPE provides reasonable fit to the attenuation rates observed. The event terms of the Mw6.7 and Mw6.9 events are positive but within the expected scatter from worldwide similar earthquakes. The intra-event standard deviations are higher than the intraevent variability of the model, which is partly related to the poorly constrained $V_{S30}$ proxys.

The Pedernales earthquake has produced a large sequence of aftershocks, with at least 9 events with magnitude higher or equal to 6.0 recorded to date. As the coast is close to the trench and the slab dip is shallow, important cities are located at short distances (20-30km) and magnitudes down to 6.0 must be included in seismic hazard studies. The next step will be to constitute a strong motion interface database and test the GMPEs with more quantitative methods (e.g. Delavaud et al. 2009, Beauval et al. 2012). On-site measurements of velocity using geophysical techniques have begun and are planed for all RENAC sites. In a year or two hopefully, the site conditions of the stations will be much better known.

**Data and resources**

The accelerometric dataset was recorded by the National Accelerometric Network of Ecuador (RENAC) maintained by the Geophysical Institute, Escuela Politécnica Nacional, Quito, and by the OCP network (Oleoducto de Crudos Pesados). The Global Centroid Moment Tensor Project database was searched using www.globalcmt.org/CMTsearch.html (last accessed 5 August 2016). The OpenQuake Ground Motion Toolkit is available online (https://github.com/GEMScienceTools/gmpe-smtk, last accessed 5 August 2016). The programs developed by D. Boore to calculate fault-to-station distances are available online (http://www.daveboore.com/software_online.html, last accessed 5 August 2016). The global $V_{S30}$ Map Server was searched using http://earthquake.usgs.gov/hazards/apps/vs30/ ((last accessed 5 August 2016).

**Acknowledgments**

This work was supported by the Instituto Geofísico, Escuela Politécnica Nacional, Quito, by the Institut de Recherche pour le Développement (IRD), CNRS-INSU, and by the Agence Nationale de la Recherche through the project REMAKE (grant ANR-15-CE04-004). On the Ecuadorian side, additional support was available from the Secretaría Nacional de Educación Superior, Ciencia y Tecnología SENESCYT (LAE-5 y Proyecto PIN_08-EPNGEO-00001). This work has been carried out in the frame of the Joint International Laboratory "Seismes & Volcans dans les Andes du Nord" (IRD LMI SVAN). Previous funding from the Agence Nationale de la Recherche of France is acknowledged (grant ANR-07-BLAN-0143-01). We also acknowledge the OCP for the use of their accelerometric data. We thank Dave Boore for sharing his programs to calculate fault-to-station distances, and Gonzalo Montava for sharing his matlab script to predict ground motions with the new Chilean GMPE. We are also grateful to Graeme Weatherhill and the Global Earthquake Model (GEM) Modeling Facility for constant support on OpenQuake. At last, we would like to acknowledge Pierre-Yves Bard for fruitful interactions and discussions.

**Tables**

Table 1. Finite fault parameters used in the present study for the M7.8 2016 Pedernales earthquake and its two largest aftershocks M6.9 and M6.7

| Date | Hour UTC | Hypocenter latitude | Hypocenter longitude | Hypocenter depth (km) | Fault strike | Dip angle | Fault length (km) | Fault width (km) | $Mw^*$ GCMT |
|---|---|---|---|---|---|---|---|---|---|
| 2016/04/16 | 23h58 | $0.35^£$ | $80.17^£$ | $17^£$ | $26.5^£$ | $23^£$ | $110^£$ | $60^£$ | 7.8 |
| 2016/05/18 | 07h57 | $0.43387^\#$ | $-80.00961^\#$ | $17^\#$ | $29^\#$ | $26^\#$ | $28^§$ | $30^§$ | 6.7 |
| 2016/05/18 | 16h46 | $0.47301^\#$ | $-79.81545^\#$ | $21^\#$ | $47^\#$ | $25^\#$ | $36^§$ | $34^§$ | 6.9 |

$^£$ deduced from Nocquet et al. (2016)

$^\#$ determined by Geophysical Institute in Quito (dip and strike obtained with Nakano et al. 2008 method)

$^§$ determined with the scaling law for interface events in Strasser et al. (2010)

$^*$ obtained from the Global Centroid Moment Tensor Project (see Data and Resources section)

Table 2. Description of RENAC accelerometer devices

|   | Sensor | Digitizer | Full scale range | Dynamic range | Frequency response | Sample frequency |
|---|---|---|---|---|---|---|
| 1 | Guralp CMG-5TD |  | ± 4g | 127 dB (3-30Hz) | DC – 100 Hz | 100 Hz |
| 2 | Reftek 130-SMA |  | ± 4g | 112 dB at 1 Hz | DC – 500 Hz | 100 Hz |
| 3 | Kinemetrics EpiSensor FBA-EST | Kephren | ± 2g | 155 dB | DC - 200 Hz | 125 Hz or 250 Hz |

**List of figure captions**

Figure 1: Location map with fault rupture and stations. The white rectangle shows the surface projection of the Pedernales mainshock Mw7.8 (inferred from Nocquet et al. 2016). Epicenters of the mainshock and its two largest aftershocks are indicated (stars). Triangles show locations of strong-motion stations, which have recorded the mainshock and/or the aftershocks. Stations with acceleration indicated (colorbar) have recorded the mainshock. Background map produced with Google Map.

Figure 2: Pedernales earthquake Mw7.8 2016/04/16. Accelerograms recorded at 6 stations around the fault plane (see Fig. 1, East component). Latitudes of stations, maximum amplitude and rupture distance to the fault plane are indicated.

Figure 3: $V_{S30}$ reference set versus rupture distance (see section "Strong motion data"). Alternative $V_{S30}$ values based on topography are in the Electronic Supplement (Fig. S1).

Figure 4: Observed spectral amplitudes of the mainshock Mw7.8, overlaid by the Abrahamson et al. (2016) predicted amplitudes (median±σ). Total sigma is indicated with dashed lines. a) PGA, for 3 different $V_{S30}$ values, forearc/unknown coefficients used for all stations; b) spectral acceleration T=1.0s, for 3 different $V_{S30}$ values, forearc/unknown coefficients used for all stations; c) PGA, predictions for forearc sites and for backarc sites for a $V_{S30}$ of 400m/s; d) spectral acceleration T=1.0s, predictions for forearc sites and for backarc sites, for a $V_{S30}$ of 400m/s.

Figure 5: Evidence of directivity effects, at the PGA (a) and 3 seconds (b). The stations located at rupture distances lower or equal to 100km are highlighted, and their location with respect to the fault plane is indicated. Abrahamson et al. (2016) predicted amplitudes with $V_{S30}$= 400m/s (see legend of Fig. 4a).

Figure 6: Observed spectral amplitudes of the mainshock Mw7.8, at PGA and spectral acceleration T=1.0s, overlaid by four GMPE curves, Abrahamson et al. (2016), Ghofrani and Atkinson (2014), Zhao et al. (2006), and Montalva et al. (2016). Total sigma is indicated with dashed lines. Predictions for an average $V_{S30}$ of 400 m/s.

Figure 7: Observed spectral amplitudes of the mainshock Mw7.8, overlaid by the Ghofrani and Atkinson (2014) predicted amplitudes (median±σ), for PGA and spectral acceleration T=1.0s, for an average $V_{S30}$ of 400m/s.

Figure 8 : Total residuals, mainshock, Abrahamson et al. (2016) model. The residuals are binned into intervals of 20km width, and the corresponding means (squares) and standard deviations (bars) are displayed when calculated on at least 4 values. Dashed lines indicate ± total sigma (0.74). Event term is the mean of the residuals. $V_{S30}$ reference set considered (see "Section Strong Motion Data"). Abrahamson et al. (2016) generating dataset does not include records beyond 300km and the model is therefore extrapolated at these distances.

Figure 9: Aftershock 2016/05/18 16h46 Mw6.9. a) and b) Attenuation of peak acceleration and spectral accelerations at T=1.0s with distance and comparison to Abrahamson et al. (2016) GMPE for an average $V_{S30}$ of 400m/s. c) and d) Total residuals of data relative to Abrahamson et al. (2016) model ; residuals binned in 20km width interval and displayed if calculated over more than 4 observations ; dashed lines indicates ±total sigma. Abrahamson et al. (2016) generating dataset does not include records beyond 300km and the model is therefore extrapolated at these distances.

Figure 10: Aftershock 2016/05/18 7h57 Mw6.7. a) and b) Attenuation of peak acceleration and spectral accelerations at T=1.0s with distance and comparison to Abrahamson et al. (2016) GMPE for an average $V_{S30}$ of 400m/s. c) and d) Total residuals of data relative to Abrahamson et al. (2016) model ; residuals binned in 20km width interval and displayed if calculated over more than 4 observations ; dashed lines indicates ±total sigma. Abrahamson et al. (2016) generating dataset does not include records beyond 300km and the model is therefore extrapolated at these distances.

Figure 11: a) Event terms of the Pedernales mainshock and its two largest aftershocks, compared to the Abrahamson et al. (2016) inter-event standard deviation τ (0.43); b) intraevent standard deviation for Pedernales mainshock and its two largest aftershocks compared to the Abrahamson et al. (2016) intra-event standard deviation, ϕ (0.6). Recordings at distances larger than 300km are not included. Results for PGA are indicated at the frequency 50Hz.

**Figures**

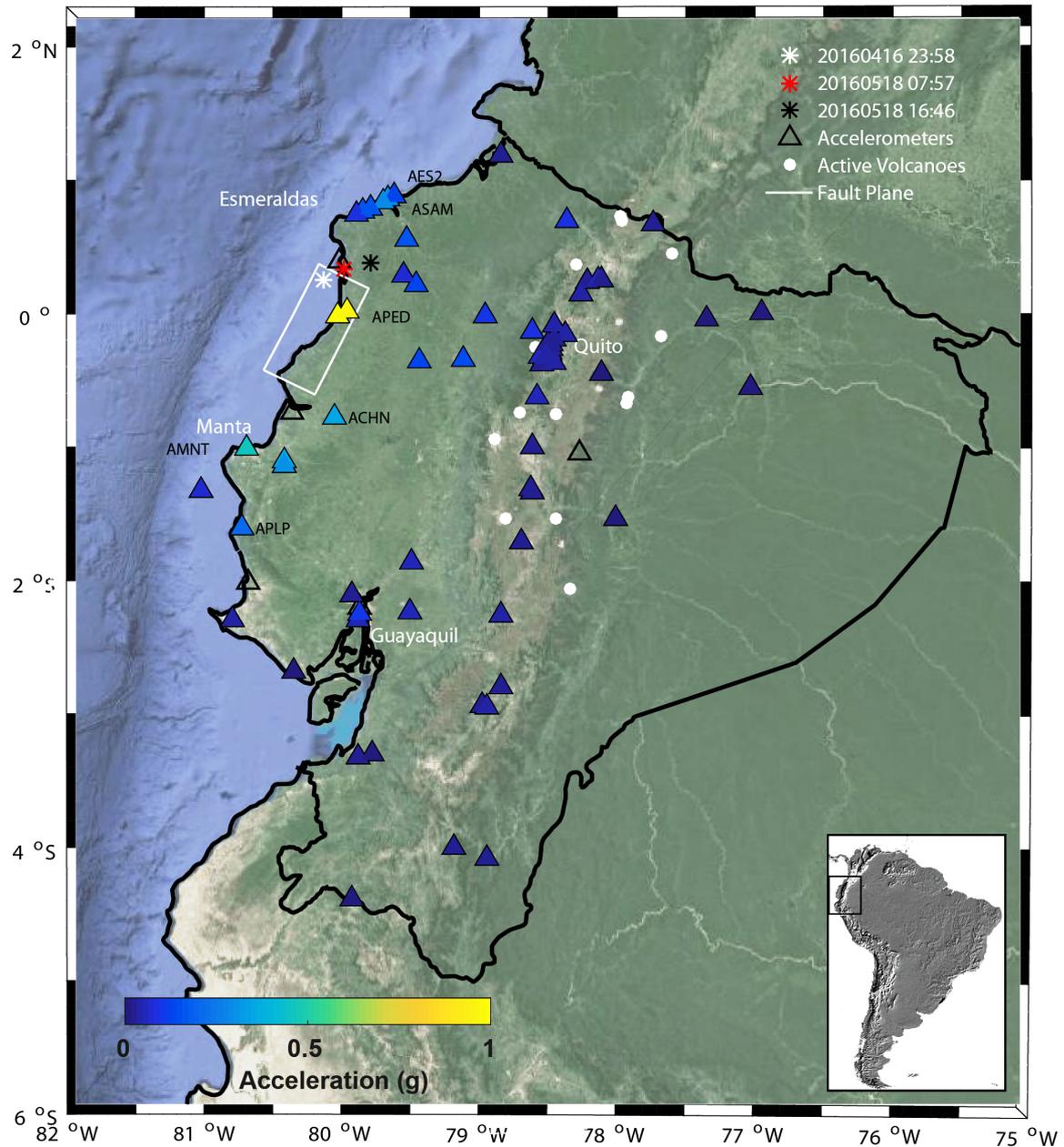

Figure 2: Location map with fault rupture and stations. The white rectangle shows the surface projection of the Pedernales mainshock Mw7.8 (inferred from Nocquet et al. 2016). Epicenters of the mainshock and its two largest aftershocks are indicated (stars). Triangles show locations of strong-motion stations, which have recorded the mainshock and/or the aftershocks. Stations with acceleration indicated (colorbar) have recorded the mainshock. Background map produced with Google Map.

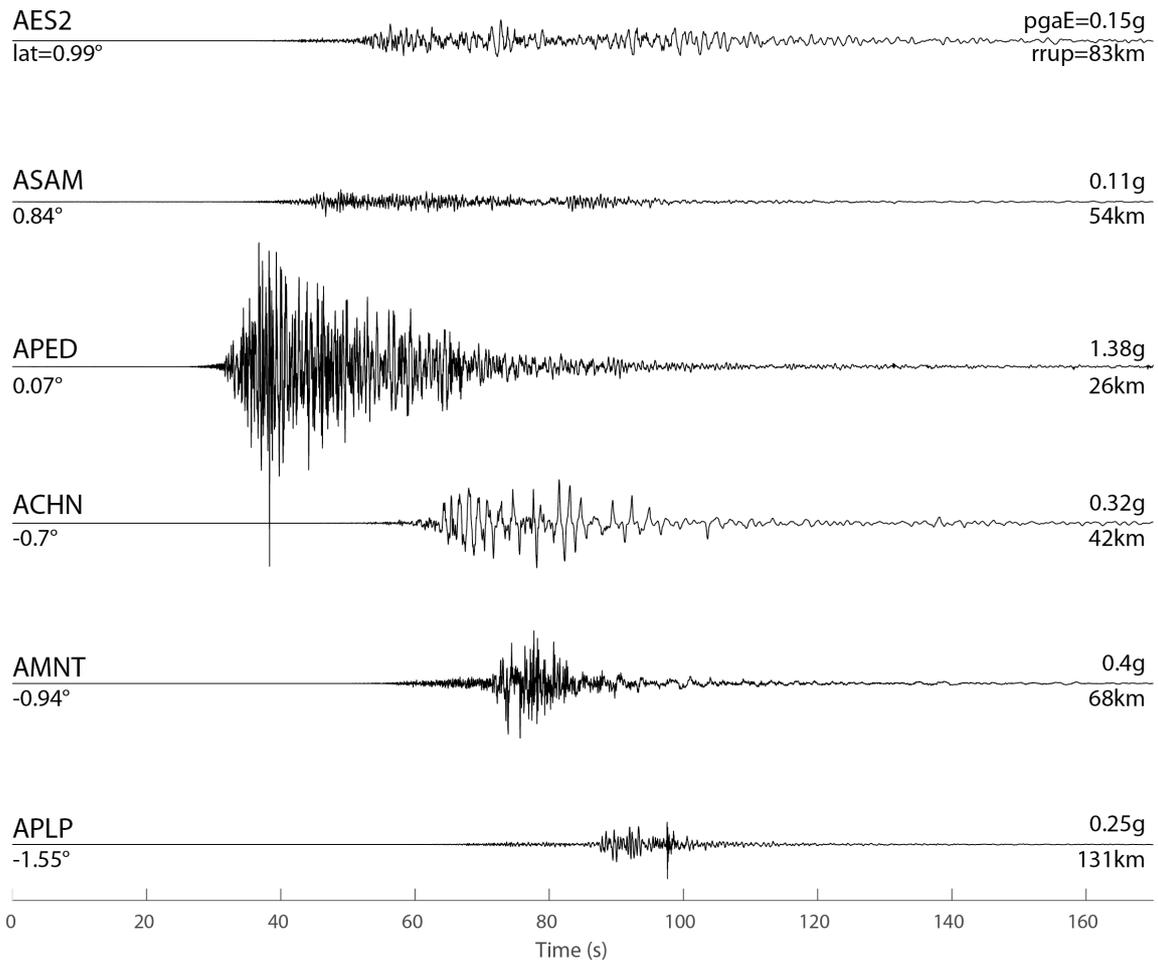

Figure 2: Pedernales earthquake Mw7.8 2016/04/16. Accelerograms recorded at 6 stations around the fault plane (see Fig. 1, East component). Latitudes of stations, maximum amplitude and rupture distance to the fault plane are indicated.

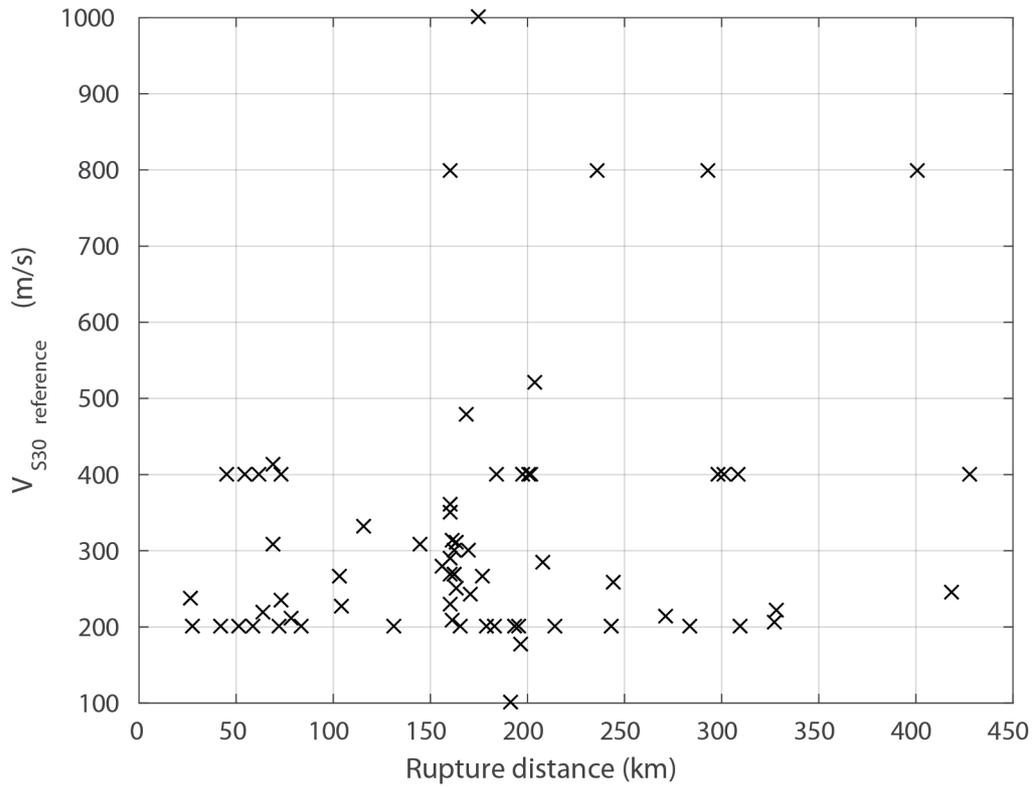

Figure 3: $V_{S30}$ reference set versus rupture distance (see section "Strong motion data"). Alternative $V_{S30}$ values based on topography are in the Electronic Supplement (Fig. S1).

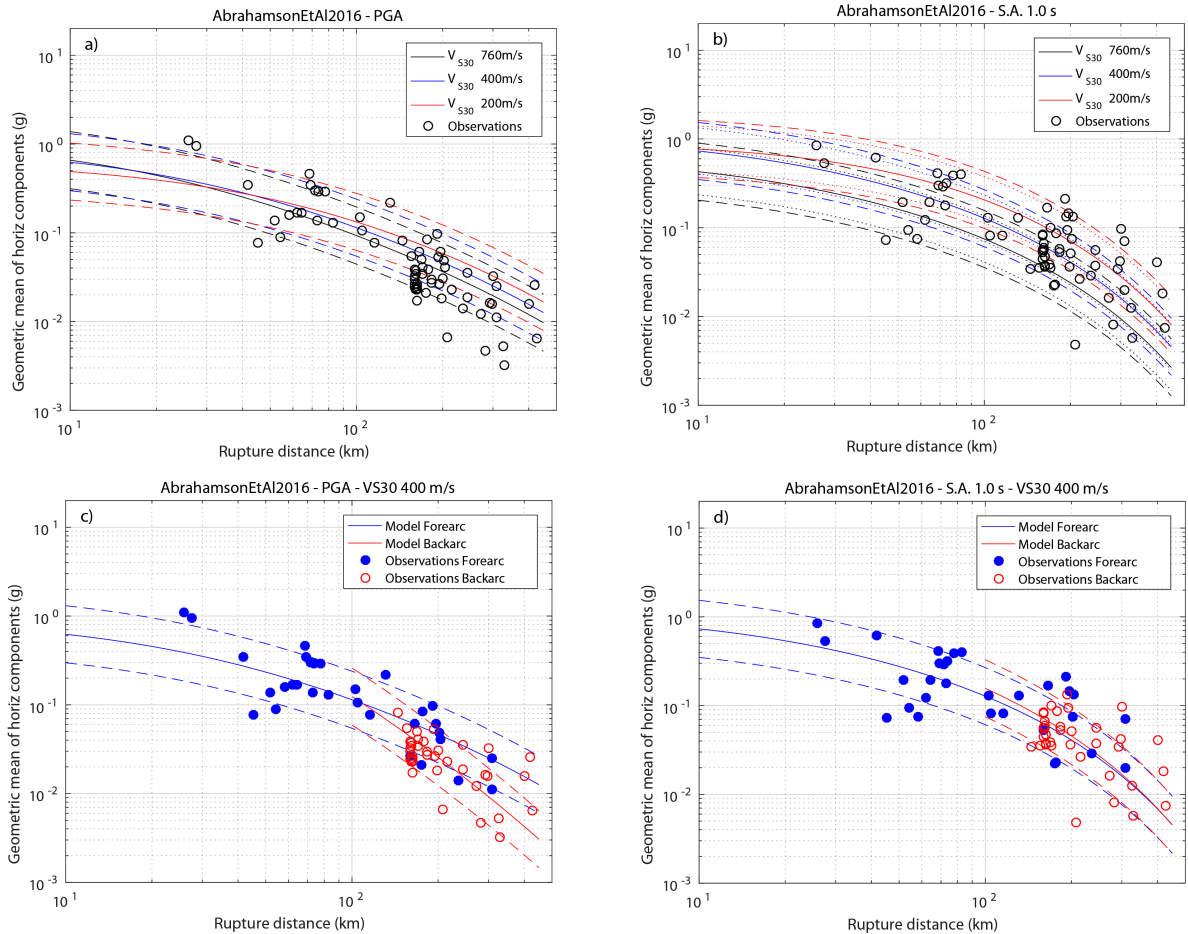

Figure 4: Observed spectral amplitudes of the mainshock Mw7.8, overlaid by the Abrahamson et al. (2016) predicted amplitudes (median±σ). Total sigma is indicated with dashed lines. a) PGA, for 3 different $V_{S30}$ values, forearc/unknown coefficients used for all stations; b) spectral acceleration T=1.0s, for 3 different $V_{S30}$ values, forearc/unknown coefficients used for all stations; c) PGA, predictions for forearc sites and for backarc sites for a $V_{S30}$ of 400m/s; d) spectral acceleration T=1.0s, predictions for forearc sites and for backarc sites, for a $V_{S30}$ of 400m/s.

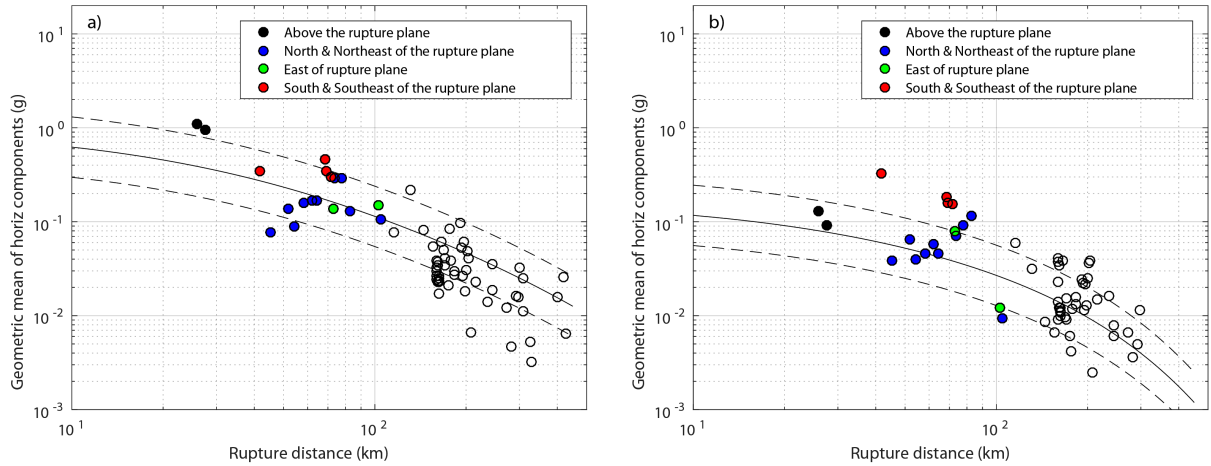

Figure 5: Evidence of directivity effects, at the PGA (a) and 3 seconds (b). The stations located at rupture distances lower or equal to 100km are highlighted, and their location with respect to the fault plane is indicated. Abrahamson et al. (2016) predicted amplitudes with $V_{S30}$= 400m/s (see legend of Fig. 4a).

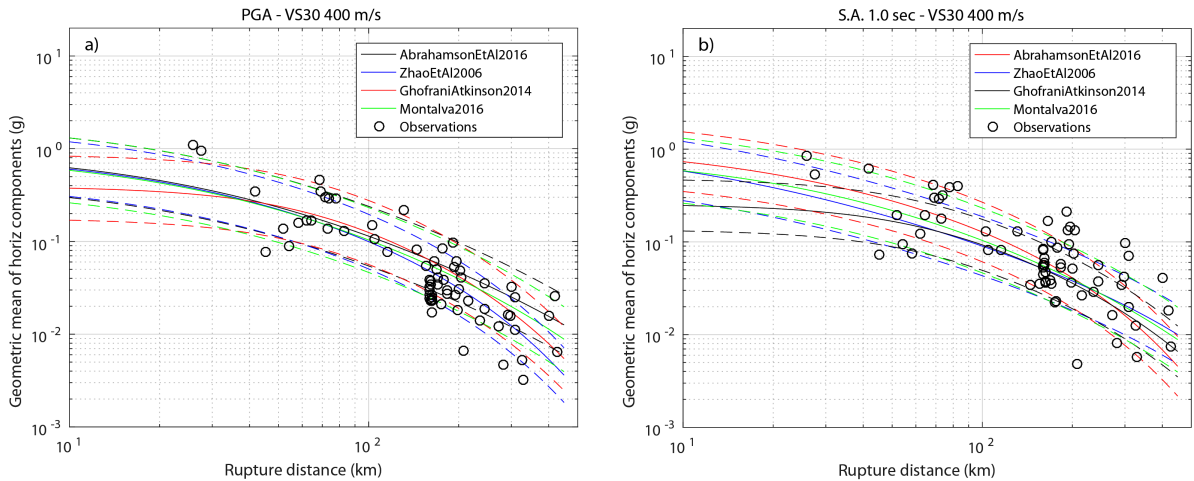

Figure 6: Observed spectral amplitudes of the mainshock Mw7.8, at PGA and spectral acceleration T=1.0s, overlaid by four GMPE curves, Abrahamson et al. (2016), Ghofrani and Atkinson (2014), Zhao et al. (2006), and Montalva et al. (2016). Total sigma is indicated with dashed lines. Predictions for an average $V_{S30}$ of 400 m/s.

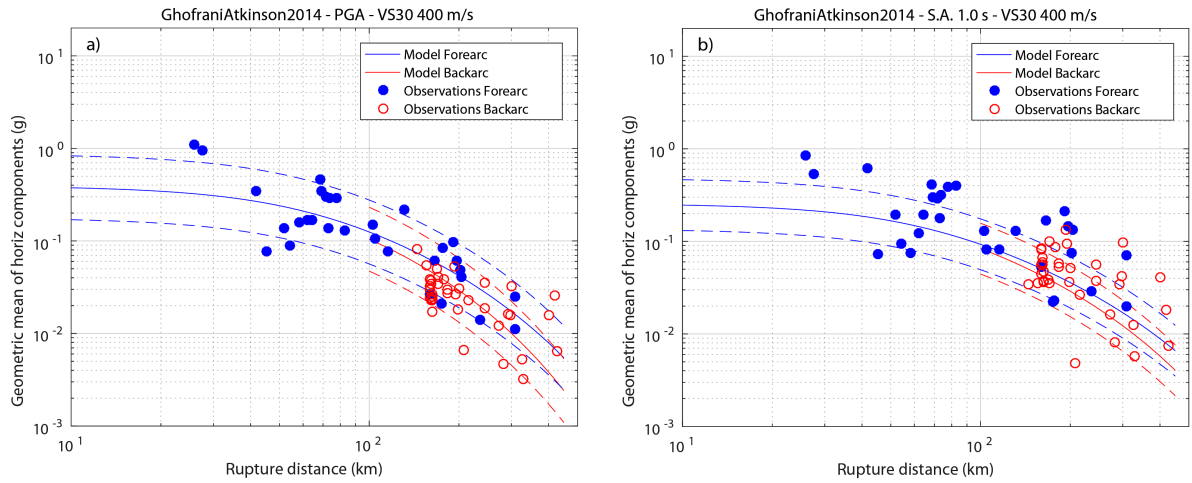

Figure 7: Observed spectral amplitudes of the mainshock Mw7.8, overlaid by the Ghofrani and Atkinson (2014) predicted amplitudes (median±σ), for PGA and spectral acceleration T=1.0s, for an average $V_{S30}$ of 400m/s.

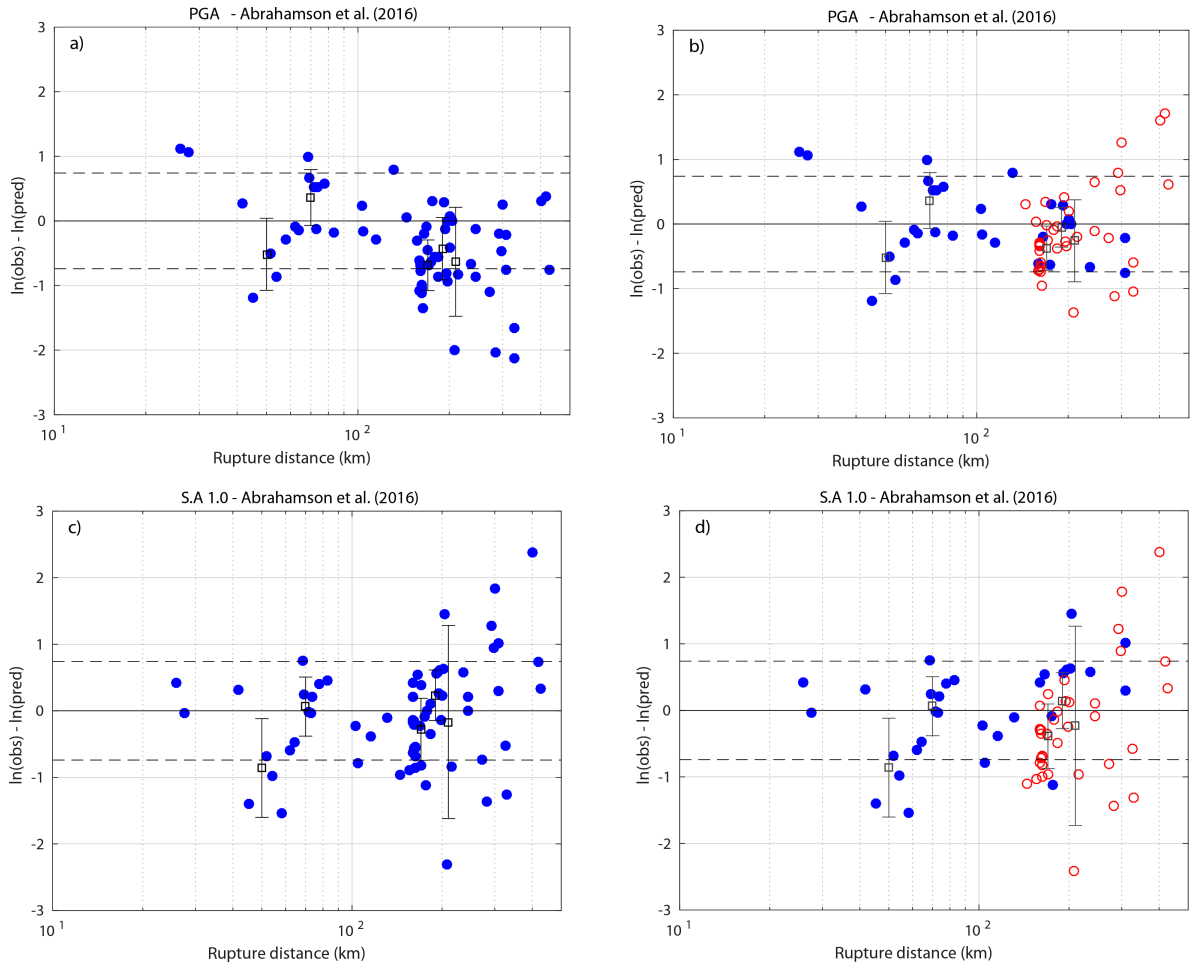

Figure 8 : Total residuals, mainshock, Abrahamson et al. (2016) model. The residuals are binned into intervals of 20km width, and the corresponding means (squares) and standard deviations (bars) are displayed when calculated on at least 4 values. Dashed lines indicate ± total sigma (0.74). Event term is the mean of the residuals. $V_{S30}$ reference set considered (see "Section Strong Motion Data"). Abrahamson et al. (2016) generating dataset does not include records beyond 300km and the model is therefore extrapolated at these distances.

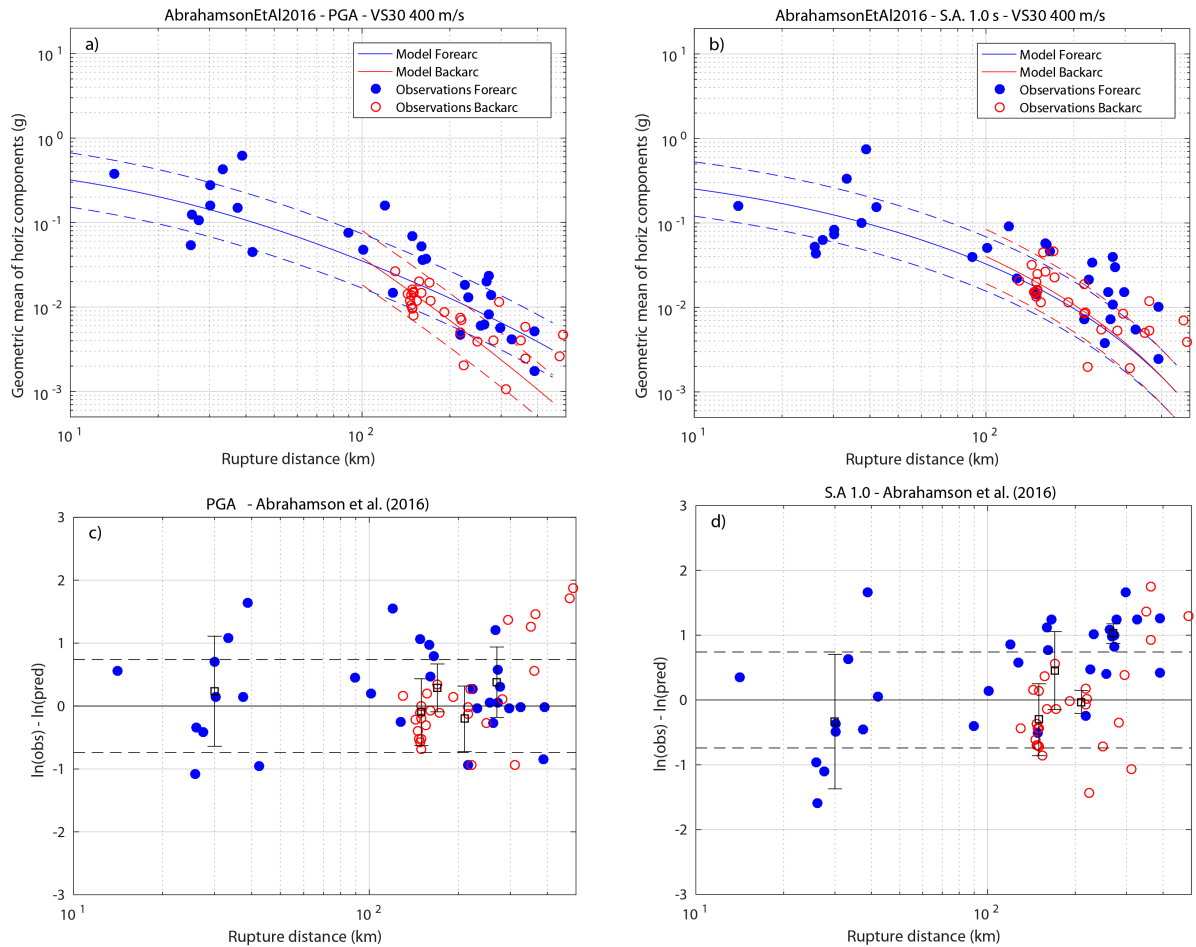

Figure 9: Aftershock 2016/05/18 16h46 Mw6.9. a) and b) Attenuation of peak acceleration and spectral accelerations at T=1.0s with distance and comparison to Abrahamson et al. (2016) GMPE for an average $V_{S30}$ of 400m/s. c) and d) Total residuals of data relative to Abrahamson et al. (2016) model ; residuals binned in 20km width interval and displayed if calculated over more than 4 observations ; dashed lines indicates ±total sigma. Abrahamson et al. (2016) generating dataset does not include records beyond 300km and the model is therefore extrapolated at these distances.

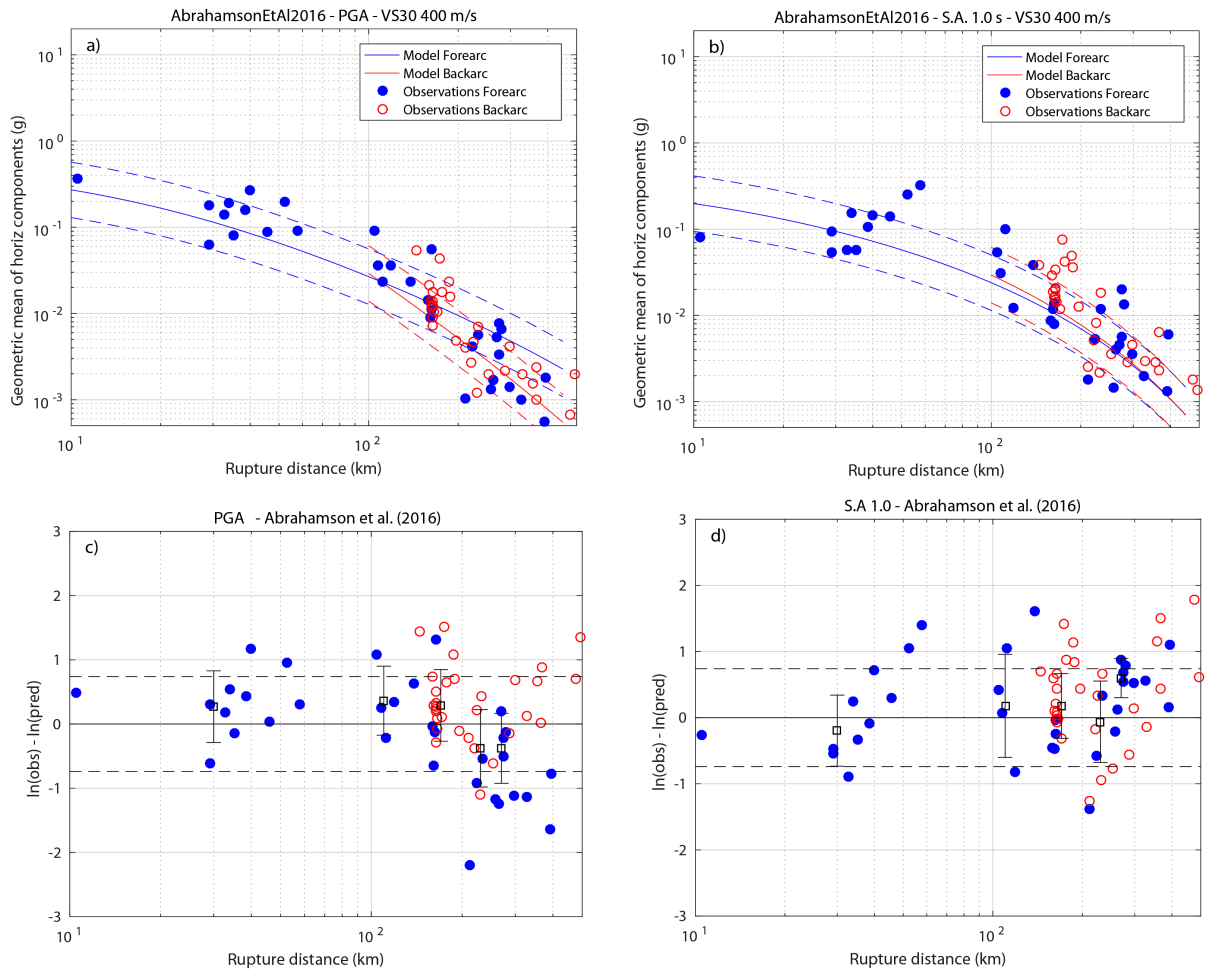

Figure 10: Aftershock 2016/05/18 7h57 Mw6.7. a) and b) Attenuation of peak acceleration and spectral accelerations at T=1.0s with distance and comparison to Abrahamson et al. (2016) GMPE for an average $V_{S30}$ of 400m/s. c) and d) Total residuals of data relative to Abrahamson et al. (2016) model ; residuals binned in 20km width interval and displayed if calculated over more than 4 observations ; dashed lines indicates ±total sigma. Abrahamson et al. (2016) generating dataset does not include records beyond 300km and the model is therefore extrapolated at these distances.

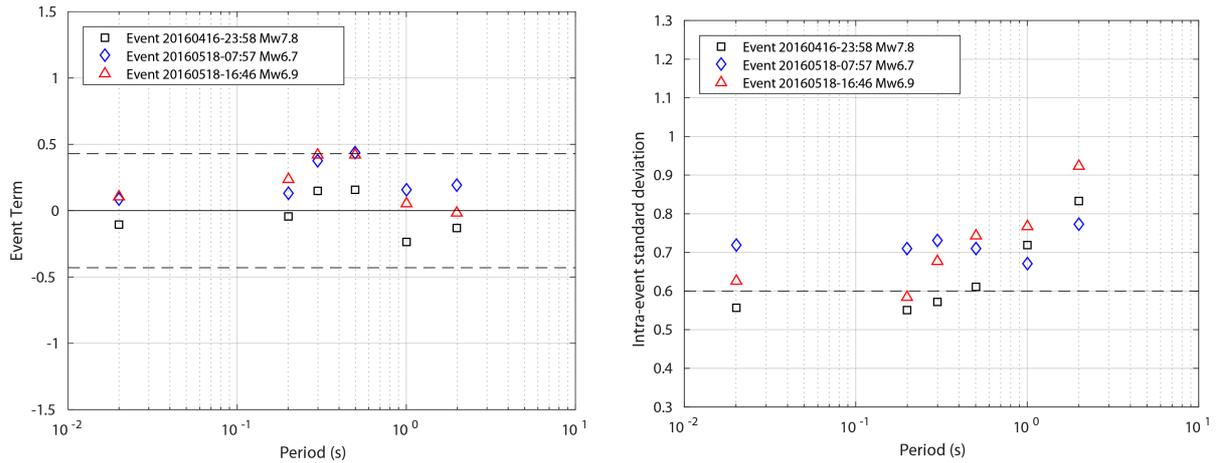

Figure 11: a) Event terms of the Pedernales mainshock and its two largest aftershocks, compared to the Abrahamson et al. (2016) inter-event standard deviation τ (0.43); b) intraevent standard deviation for Pedernales mainshock and its two largest aftershocks compared to the Abrahamson et al. (2016) intra-event standard deviation, φ (0.6). Recordings at distances larger than 300km are not included. Results for PGA are indicated at the frequency 50Hz.